# A New Approach to the Synthesis of Nanocrystal Conjugated Polymer Composites

A.A.R. Watt*, H. Rubinsztein-Dunlop & P. Meredith

Soft Condensed Matter Physics Group, School of Physical Sciences, University of Queensland, Queensland 4072, Australia.

Abstract

A novel one pot process has been developed for the preparation of PbS nanocrystals in the conjugated polymer poly 2-methoxy,5-(2 -ethyl-hexyloxy-p-phenylenevinylene) (MEH-PPV). Current techniques for making such composite materials rely upon synthesizing the nanocrystals and conducting polymer separately, and subsequently mixing them. This multi-step technique has two serious drawbacks: templating surfactant must be removed before mixing, and co-solvent incompatibility causes aggregation. In our method, we eliminate the need for an initial surfactant by using the conducting polymer to terminate and template nanocrystal growth. Additionally, the final product is soluble in a single solvent. We present materials analysis which shows PbS nanocrystals can be grown directly in a conducting polymer, the resulting composite is highly ordered and nanocrystal size can be controlled.

## 1. Introduction

Several groups have blended nanocrystals with conjugated polymers for use in optoelectronic devices [1, 2]. Recently, we developed a new synthesis for lead sulfide (PbS) nanocrystals in the conjugated polymer poly 2-methoxy,5-(2 -ethyl-hexyloxy-p-phenylenevinylene) (MEH-PPV) [3].

The current techniques for making nanocrystal: conjugated polymer composite materials rely upon synthesizing nanocrystals separately, and then mixing them with the conjugated polymer [4]. This approach has two shortcomings: firstly, a surfactant must be used to control nanocrystal size and shape. Some of the surfactant becomes incorporated into the final nanocrystal and conjugated polymer mix, which inhibits efficient charge transfer. Secondly, the mixing approach requires the use of co-solvents, which can adversely affect nanocrystal solubility and polymer chain orientation.

The major advantage of the new method we describe in this paper is that it eliminates the need for an initial surfactant to terminate nanocrystal growth, and also eliminates the need for subsequent transfer to the conjugated polymer. A similar method has been proposed by Milliron et al. [5] which utilizes an electro-active surfactant. Although our method does not allow tight control of nanocrystal size distribution, it does allow more intimate contact between nanocrystal and the conjugated polymer backbone, which we believe will enhance electronic coupling between the two components and hence improve charge transfer in the system. It is also a significantly less complicated synthetic route.

Our novel approach uses the conjugated polymer MEH-PPV to control the nanocrystal growth and passivate surface states. MEH-PPV has a high hole mobility and low electron mobility [6]. This relative imbalance limits the performance of any optoelectronic device based upon the material. Nanocrystals, by acting as a percolated high mobility pathway for electrons, offsets this imbalance [7]. It is thought that photoexcited charge separation occurs at the nanocrystal-polymer interface [8]. Hence, the conjugated polymer acts as a colloidal template, and also as the continuous conductive matrix through which photogenerated charges are transferred to the external circuit. We choose lead sulphide (PbS) as the inorganic material because, in the quantum regime, it has a broad band absorption [9]. Additionally, the electrons and holes are equally confined in PbS nanocrystals, [9] and they been shown to exhibit long excited state lifetime [10].

---

* Corresponding author, Tel: +61 7 3365 1245; fax: +61 7 3365 1242; E-mail: watt@physics.uq.edu.au

In this paper we discuss new results from our most refined synthesis to date. In particular we show how polymer molecular weight and solvent ratios can be used to control nanocrystal size.

## 2. Experimental

The nanocrystal: conjugated polymer composite was prepared as follows: A sulphur precursor solution was prepared by dissolving 0.08g of sulphur flakes in 10ml of toluene. The mixture was stirred and degassed with argon for 1 hour. In a typical synthesis, 9ml of toluene, 0.01g of 80,000 Daltons average molecular weight MEH-PPV, 3ml of di-methylsulfoxide DMSO and 0.1g of lead acetate were mixed and degassed with argon at 100 ºC for 2 hours in a 25 ml three-neck flask connected to a Liebig condenser. All materials where purchased from Sigma Aldrich and used without further purification. The resultant solution was bright orange in colour with no precipitate or solvent separation. With the solution at 160 ºC, 1ml of the sulphur precursor was injected. The reaction took approximately 15 minutes to reach completion upon which a brown solution resulted. The product was cleaned to remove excess lead or sulphur ions, DMSO and low molecular weight MEH-PPV by adding the minimum amount of anhydrous methanol to cause precipitation of the composite material. The sample was centrifuged and the supernatant removed. The precipitate was then redissolved in the desired solvent (for example toluene or chlorobenzene). Through the reaction 0.2ml samples where taken every three minutes and the reaction halted by injecting into toluene at ambient temperature. This synthesis is referred to as 3:1, Toluene: DMSO henceforth.

The synthesis was repeated twice with a single variation each time. The first used 22,000 Daltons MEH-PPV instead of 80,000 Daltons and the second used 8ml of toluene and 4ml of di-methylsulfoxide DMSO (refered to as 2:1, Toluene: DMSO).

Transmission electron microscopy (TEM) was carried out using a Tecnai 20 Microscope. Samples where prepared by taking the cleaned product, diluting it and placing a drop on an ultra thin carbon coated copper grid (Ted Pella) with the Formvar removed. A Perkin-Elmer λ40 UV-Visible Spectrophotometer was used to obtain absorption spectra of spun-cast films.

## 3. Results and Discussion

TEM was used to gain an understanding of the nanocrystal growth and quality. Figure 1 shows that nanocrystals are formed and they are non-aggregated with an average size of 4nm (±2nm). Figure 2 shows the crystal lattice of an individual nanocrystal and demonstrates a high degree of crystallinity. Figure 3 shows a 2μm selected area diffraction pattern of a field of nanocrystals. The diffraction corresponds to the lattice parameter and pattern of cubic PbS looking down the [1,1,1] zone axis. Usually samples prepared from colloidal solutions display only circular poly-crystalline electron diffraction patterns. The diffraction pattern in figure 3 would tend to indicate a low degree of orientational anisotropy at the ensemble level. These are similar results to those reported by Berman et al. [11] who showed that an ordered array of nanocrystals could form in a polymer matrix.

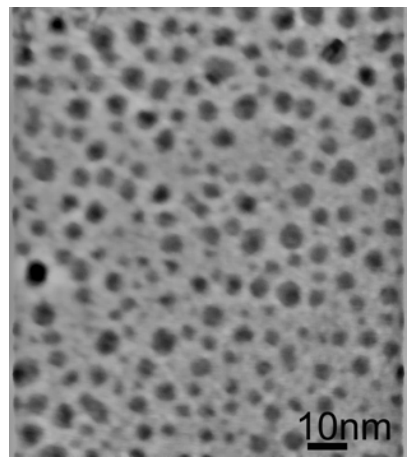

Fig. 1. TEM image of a field of nanocrystals prepared with 3:1, Toluene: DMSO and 80, 000 Daltons MEH-PPV.

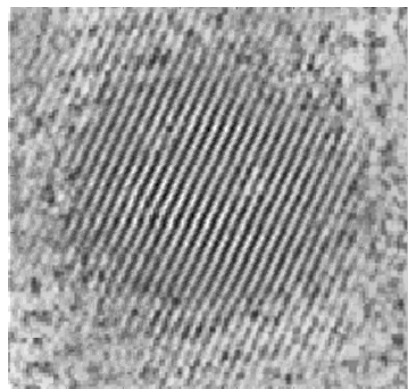

Fig.2. High resolution TEM image of the lattice planes in a single nanocrystal prepared with 3:1, Toluene: DMSO and 80, 000 Daltons MEH-PPV.

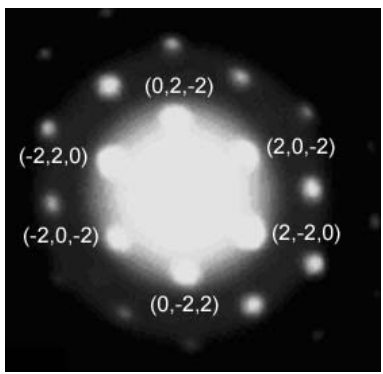

Fig. 3. Selected area diffraction image using a 2μm aperature on a field of nanocrystals prepared with 3:1, Toluene: DMSO and 80, 000 Daltons MEH-PPV.

MEH-PPV has an absorption edge at around 560nm, Figure 4 shows how the absorption changes as PbS nanocrystals assemble through the reaction. This results in an extension of the absorption into the near IR. The absorption edge corresponds to theoretical predictions for PbS nanocrystals of between 4 and 6 nm.

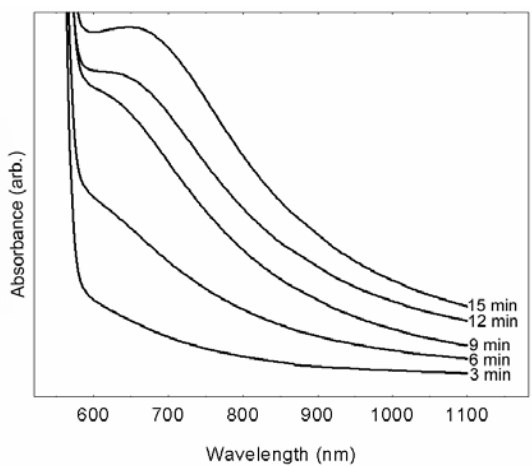

Fig. 4. Change in absorption as reaction proceeds, prepared with 3:1, Toluene: DMSO and 80, 000 Daltons MEH-PPV.

With respect to kinetics, we find that nanocrystal growth is dependent on reaction temperature, reaction time, polymer chain length and polymer solvation. In a standard nanocrystal synthesis, growth control is derived from a combination of electrostatic effects from the surfactant functional groups (eg phosphine), and the steric effects of the long surfactant chain (typically C18 to C24). MEH-PPV has no charged functional groups which could electrostatically control nanocrystal growth. Therefore we believe that growth is predominantly influenced by steric effects of the long chain MEH-PPV. Not surprisingly, the rate of reaction is greater at elevated temperatures due to increased solubility and reactivity of the precursors. Figure 5 shows that using 22,000 Daltons MEH-PPV produces larger nanocrystals. The MEH-PPV used in all our reactions is unpurified and contains a spread of molecular weights; this we believe influences the size dispersity of nanocrystals.

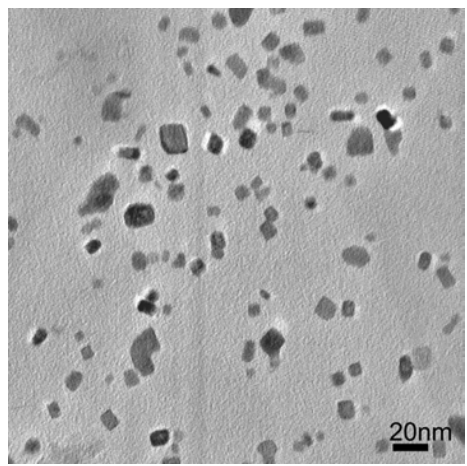

Fig. 5. TEM image of a field of nanocrystals, prepared with 3:1, Toluene: DMSO and 22, 000 Dalton MEH-PPV.

If the ratio of Toluene to DMSO used in the reaction is changed from 3:1 to 2:1 we find that the reaction kinetics change as shown by the time evolution absorption graph in figure 6. Note the red shift of the nanocrystal absorption edge, we attribute this to a change in mean nanocrystal size [9]. Further microscopy studies are underway to determine the size of nanocrystals at each stage of the reaction.

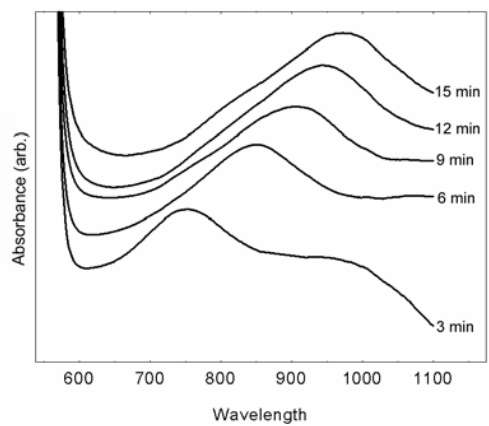

Fig. 6. Change in absorption as reaction proceeds, prepared with 2:1, Toluene: DMSO and 80, 000 Daltons MEH-PPV.

## 4. Conclusions

In conclusion we have demonstrated that it is possible to make nanocrystals in a conjugated polymer by a simple single step process without the need for additional surfactants. The nanocrystals self assemble, are highly crystalline and have absorption characteristics as predicted by theory. Initial results tend to indicate that we can tune the nanocrystal size by a combination of solvent ratios and time.

Although this method seems particularly suited to PbS in MEH-PPV, it could potentially be applied to other sorts of nanocrystals e.g. CdSe and other conjugated polymers. Further work is underway to understand the complex dynamics of nanocrystal growth using different polymer molecular weights, purifying the polymer to yield a narrower distribution of molecular weights, and using other solvent systems in a bid to control nanocrystal size and dispersity. We envisage device applications in photovoltaics in particular.

It is worthy of note that we have recently demonstrated a solar cell with ~1% power conversion efficiency. These results are to be presented elsewhere.

## Acknowledgments

The work was funded by the Australian Research Council. TEM was performed at the University of Queensland Centre for Microscopy and Microanalysis.